\documentclass[conference]{IEEEtran}
\IEEEoverridecommandlockouts
\usepackage{cite}
\usepackage{amsmath,amssymb,amsfonts}
\usepackage{algorithmic}
\usepackage{graphicx}
\usepackage{pifont}
\newcommand{\cmark}{\ding{51}}%
\newcommand{\xmark}{\ding{55}}%
\usepackage{textcomp}
\usepackage{mathtools}
\usepackage{amsthm}
\usepackage{bm}
\usepackage{bbm}
\usepackage{url}
\usepackage{multirow}
\usepackage{subcaption}
\usepackage{xcolor}
\def\BibTeX{{\rm B\kern-.05em{\sc i\kern-.025em b}\kern-.08em
    T\kern-.1667em\lower.7ex\hbox{E}\kern-.125emX}}
\begin{document}

\title{Multi-modal Spatial Clustering for Spatial Transcriptomics Utilizing High-resolution Histology Images
\thanks{This study is funded by National Science Foundataion (NSF)}
}

\author{\IEEEauthorblockN{1\textsuperscript{st} Bingjun Li}
\IEEEauthorblockA{\textit{School of Computing} \\
\textit{University of Connecticut}\\
Storrs, USA \\
bingjun.li@uconn.edu}
\and
\IEEEauthorblockN{2\textsuperscript{nd} Mostafa Karami}
\IEEEauthorblockA{\textit{School of Computing} \\
\textit{University of Connecticut}\\
Storrs, USA \\
mostafa.karami@uconn.edu}
\and
\IEEEauthorblockN{3\textsuperscript{nd} Masum Shah Junayed}
\IEEEauthorblockA{\textit{School of Computing} \\
\textit{University of Connecticut}\\
Storrs, USA \\
masumshah.junayed@uconn.edu}
\and
\IEEEauthorblockN{4\textsuperscript{rd} Sheida Nabavi}
\IEEEauthorblockA{\textit{School of Computing} \\
\textit{University of Connecticut}\\
Storrs, USA \\
sheida.nabavi@uconn.edu}
}
\maketitle

\begin{abstract}
Understanding the intricate cellular environment within biological tissues is crucial for uncovering insights into complex biological functions. While single-cell RNA sequencing has significantly enhanced our understanding of cellular states, it lacks the spatial context necessary to fully comprehend the cellular environment. Spatial transcriptomics (ST) addresses this limitation by enabling transcriptome-wide gene expression profiling while preserving spatial context. One of the principal challenges in ST data analysis is spatial clustering, which reveals spatial domains based on the spots within a tissue. Modern ST sequencing procedures typically include a high-resolution histology image, which has been shown in previous studies to be closely connected to gene expression profiles. However, current spatial clustering methods often fail to fully integrate high-resolution histology image features with gene expression data, limiting their ability to capture critical spatial and cellular interactions.

In this study, we propose the spatial transcriptomics multi-modal clustering (stMMC) model, a novel contrastive learning-based deep learning approach that integrates gene expression data with histology image features through a multi-modal parallel graph autoencoder. We tested stMMC against four state-of-the-art baseline models: Leiden, GraphST, SpaGCN, and stLearn on two public ST datasets with 13 sample slices in total. The experiments demonstrated that stMMC outperforms all the baseline models in terms of ARI and NMI. An ablation study further validated the contributions of contrastive learning and the incorporation of histology image features.
\end{abstract}


\section{Introduction}
Biological tissue samples contain highly complex cellular processes, which are shaped by cell distribution patterns, cell types, cell states, composition, and cell-cell interactions~\cite{janiszewska2020microcosmos}. Such information is crucial for understanding tissue development, repair, and responses to external signals~\cite{greaves2012clonal,janiszewska2020microcosmos}. Single-cell RNA sequencing technology has evolved dramatically in recent years to be more efficient, accessible, and accurate, which enables researchers to obtain deep insights into cellular states and led to the discovery of new cell types~\cite{papalexi2018single,wang2024qot}. However, while single-cell sequencing provides valuable insights, the lack of crucial contextual information limits the understanding of how cells cohabit, interact, and communicate within their environments~\cite{tanay2017scaling,li2023contrastive}.

Spatial transcriptomics (ST) addresses this gap by enabling transcriptome-wide gene expression profiling while preserving spatial context\cite{vandereyken_methods_2023,asp_spatially_2020}, which enables researchers to move beyond the scope of cell clustering to find higher-order tissue structures. Spatial clustering on ST data has become a standard first step for any downstream analysis, such as tissue anatomy visualization, finding domain-dependent biomarkers, and identifying domain-specific molecular regulatory networks~\cite{yuan_benchmarking_2024,chang2022define,hu_spagcn_2021,cable2022cell}. As the amount of ST data rapidly expands, with technologies such as Visium~\cite{maynard2021transcriptome}, seqFISH+~\cite{eng2019transcriptome}, and MERFISH~\cite{chen2015spatially} becoming more accessible, there is a growing need for advanced spatial clustering methods that can handle this kind of complex data. Current ST procedures also contain a high-resolution histology image. While it has been proven in prior studies that histology image features and gene expression are closely linked~\cite{jia_thitogene_2023,pang2021leveraging}, most current analytical methods do not fully integrate spatial information and histology image with gene expression data. This limitation obstructs the clustering models from extracting critical information from the histology images, such as cell-cell interactions and spatial changes in cell states.

Given these challenges, finding patterns in gene expression profiles with spatial and image context remains one of the significant challenges in spatial transcriptomics analysis. To address this, we propose a novel contrastive learning-based deep learning model named the spatial transcriptomics multi-modal clustering (stMMC) model. Our model integrates both gene expression and histology image features through a parallel graph autoencoder, leveraging contrastive learning to regulate feature extraction for each modality. This approach aims to provide a more accurate and comprehensive analysis of spatial transcriptomics data.

\begin{figure*}[ht]
\begin{center}
\centerline{\includegraphics[width=\textwidth]{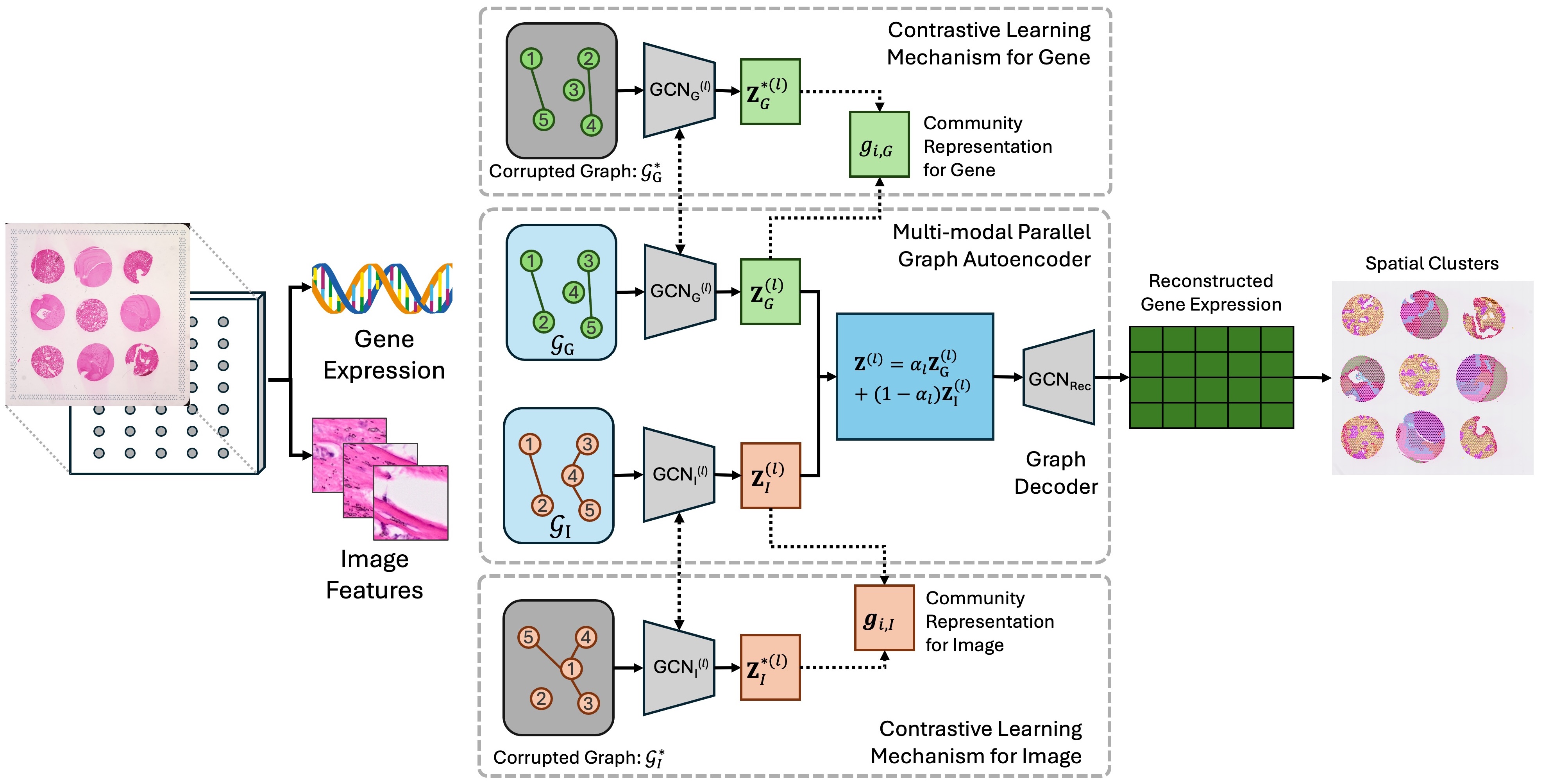}}
\caption{The overall structure of the proposed model, stMMC is plotted here, where trapezoids represent the GCN layer, and rectangles represent extracted features. Dashed lines with double arrowheads represent that both GCNs share the same weight. stMMC takes two data modalities and passes them through the multi-modal parallel graph autoencoder (MPGA), where each modality is regulated by a contrastive learning mechanism. The detailed process of contrastive learning is shown in Figure \ref{fig:contrastive_learning}. The MPGA reconstructs a refined gene expression data, which is then used for spatial clustering.}
\label{fig:overall_structure}
\end{center}
\vspace{-20pt}
\end{figure*}

Our contributions of this study are: i) we propose stMMC, a novel multi-modal contrastive learning-based clustering method using high-resolution histology images for spatial clustering in spatial transcriptomics data; ii) we demonstrate that combining gene expression data and learned histology image features improves the spatial clustering performance and paves a new way for future research; iii) we conduct experiments to show the proposed method achieves superior performance on benchmark datasets.

\begin{figure*}[ht]
\begin{center}
\centerline{\includegraphics[width=0.8\textwidth]{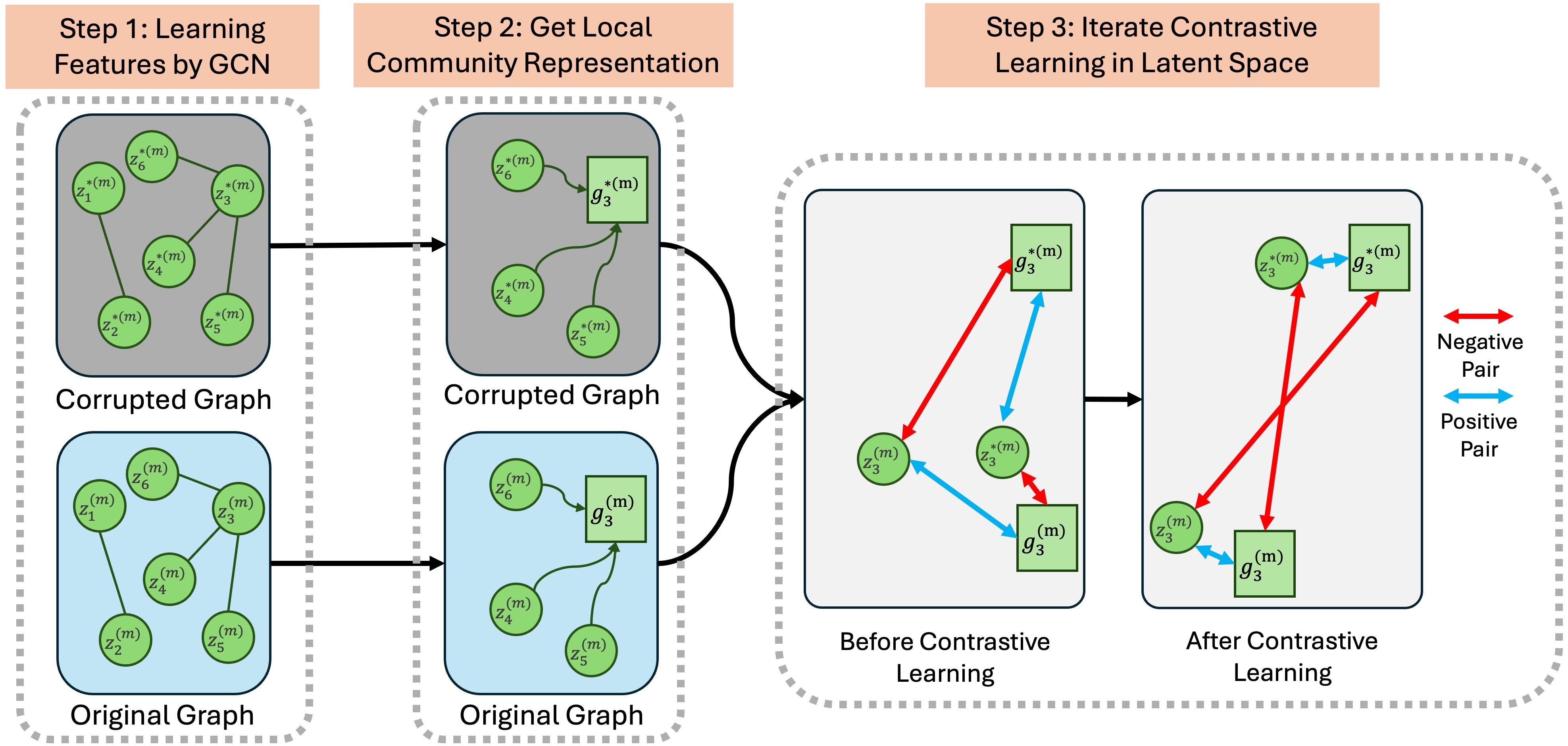}}
\caption{The detailed process of contrastive learning mechanism for a random spot on any modality is plotted here, where the top row is the corrupted graph, the bottom row is the original graph, and there is three steps of the contrastive learning mechanism: 1) obtaining the learned spot feature by GCN; 2) computing the original local community representation and the corrupted one; 3) assigning positive pairs to the learned features and the community representation from the same graph, and negative pair to the learned feature and the community representation from different graphs. The positive pair is shown in blue and negative pair is shown in red.}
\label{fig:contrastive_learning}
\end{center}
\vspace{-20pt}
\end{figure*}

\section{Related Works}
As ST sequencing has emerged as a powerful tool for mapping gene expression profiles with spatial context, researchers have proposed several methods to tackle the spatial clustering problem~\cite{luecken_benchmarking_2022}. Neural network has emerged as a powerful feature extractor in recent years~\cite{he2023robust}. These methods can be divided into three major categories based on how they extract information from the gene expression: 1) conventional-based methods; 2) neural-network-based methods using only location information; and 3) neural-network-based methods using both location information and image features.

Zhao et al. proposed BayesSpace, which is a Bayesian statistical model that utilizes spatial neighborhood information to refine the expression data and clustering on a low-dimension representation of the data~\cite{zhao_spatial_2021}. Li et al. proposed BASS (Bayesian Analytics for Spatial Segmentation), which is able to do multi-sample and multi-scale analysis and utilizes a Bayesian hierarchical modeling framework for spatial clustering~\cite{li_bass_2022}. Leiden is a widely used clustering model on graph-structured data that is not designed for ST data~\cite{traag2019louvain}. It is an improved version of the Louvain model and works by iteratively optimizing the partitions to identify densely connected communities within a network. These conventional-based methods are very efficient for training and inferencing, but they do not deal well with the high dimensionality of the ST data.

Neural-network-based deep learning methods have shown great results of extracting complex features in many field~\cite{he2023robust_uncertainty}. Most neural-network-based methods represent the location information in the form of graphs and utilize some kind of graph neural network (GNN) to extract a low-dimension representation out of the gene expression data~\cite{li2021cancer,wang2021single}. Ren et al. proposed SpaceFlow, which leverages deep GNN to integrate gene expression similarity with spatial information, producing spatially consistent low-dimensional embeddings by spatial regularization, which is used for clustering~\cite{ren_identifying_2022}. Fang et al. proposed stAA, which is an adversarial variational graph autoencoder for spatial clustering that uses GNN to combine gene expression and location information and enforces the distribution of extracted features to a prior distribution through Wasserstein distance~\cite{fang_staa_2023}. Xu et al. proposed SEDR, which integrates learned lower-dimension gene expression features with the corresponding spatial information using a variational graph autoencoder~\cite{xu_unsupervised_2024}. 

All of these methods do not utilize the information of histology images and the few methods that do incorporate histology images only use tissue morphology information from the images. Hu et al. proposed SpaGCN, which uses graph convolution network (GCN) to combine gene expression data, location information, and a three-dimensional coordinate $z$, computed from the RGB values of the histology image features~\cite{hu_spagcn_2021}. Pham et al. proposed StLearn, which utilizes spatial morphological gene expression (SME)-normalized data to perform unsupervised clustering by grouping similar points into clusters~\cite{pham_robust_2023}.

\section{Methods}
As shown in Figure \ref{fig:overall_structure}, the proposed model consists of three major modules: i) the multi-modal parallel graph autoencoder (MPGA) that consists of two independent graph autoencoders (GAEs), one for each modality -- gene expression data and learned histology image patch features; ii) the contrastive learning module that regulates each GAE through a contrastive learning mechanism utilizing a corrupted graph; iii) the decoder \& cluster module that reconstructs the gene expression using a graph decoder and takes the refined gene expression data through a cluster to obtain the final spatial cluster assignments.

\subsection{Problem Formulation}
Consider a spatial transcriptomics dataset that has $N$ spots with $M$ number of gene sequencing readings, which is denoted as $\textbf{X}_{\text{G}} = \left[ \bm{x}_1^{G}, \bm{x}_2^G,...,\bm{x}_N^G\right] \in \mathbb{R}^{N \times M}$. $M=3000$ genes with the highest variance are selected as default for stMMC. For histology images, a square patch corresponding to each spot is extracted. An autoencoder pre-trained on ImageNet is then used to extract image features from these patches. The extracted image features are denoted as $\textbf{X}_{\text{I}} \in \mathbb{R}^{N \times D}$, where $D$ is the dimension of the image features.

To better utilize the location information along with gene expression data and histology image features, we generated a graph for each modality that incorporates spot relationship information from the other modality. For example, we created a graph for the gene expression modality using spot proximity information from the histology image modality and vice versa. Thus, we are able to fuse the information from different modalities before the fusion by aggregation. The graph for the gene expression modality is defined as $\mathcal{G}_{\text{G}}=(\textbf{X}_{\text{G}},V_{\text{S}},E_G$, where $\textbf{X}_{\text{G}}$ is the gene expression data, $V_S$ is the set of nodes that each represents a spot on the sample, and $E_G$ is the set of connecting edges based on the proximity between spots. The corresponding adjacency matrix is denoted as $\textbf{A}_G$, where $\textbf{A}_{ij}=1$ when spot $i$ and $j$ are close in distance and $\textbf{A}_{ij}=0$ when two spots are far away. For any spot $i$ in the $\mathcal{G}_{\text{G}}$, $K=3$ nearest spots are selected to be connected. The graph for the image feature modality is defined as $\mathcal{G}_{\text{I}}=(\textbf{X}_{\text{I}},V_{\text{S}}, E_I)$, where $\textbf{X}_{\text{I}}$ is the image features, $V_{\text{S}}$ is the set of spots on the sample and $E_I$ is the set of edges based on similarities between gene expression data. To compute the similarity edges, we used PCA to reduce the dimension of the gene expression data and then ran KNN based on the Euclidean distance among the gene expression features to find the $K=3$ nearest spots for each spot. The corresponding adjacency matrix is $\textbf{A}_I$, where $\textbf{A}_{ij}=1$ when the expression data from spots $i$ and $j$ are similar and $\textbf{A}_{ij}=0$ when the expression data from two spots are quite different. In summary, we create a unique graph for each modality that shares the same set of nodes, but with different node attributes and different sets of edges that carry information from the other modality.

\subsection{Multi-modal Parallel Graph Autoencoder}

To adequately extract information from each modality, two independent GAEs are used within the MPGA. 

\begin{align}
    \textbf{Z}^{(l)}_{\text{G}}&=\sigma(\Tilde{\textbf{A}}_G\textbf{Z}^{(l-1)}_{\text{G}}\textbf{W}^{(l-1)}_{\text{G}}+\textbf{B}^{(l-1)}_{\text{G}}), \\
    \textbf{Z}^{(l)}_{\text{I}}&=\sigma(\Tilde{\textbf{A}}_I\textbf{Z}^{(l-1)}_{\text{I}}\textbf{W}^{(l-1)}_{\text{I}}+\textbf{B}^{(l-1)}_{\text{I}}),
\end{align}
where $l$ is the number of layers in the GAE, $\Tilde{\textbf{A}}_G$ and $\Tilde{\textbf{A}}_I$ are both the normalized adjacency matrices, $\textbf{Z}^{(l)}_{\text{G}} \in \mathbb{R}^{N \times F}$ and $\textbf{Z}^{(l)}_{\text{I}} \in \mathbb{R}^{N \times F}$ are the learned features from the $l$-th layer in the gene expression GAE and image feature GAE, respectively, $F$ is the length of the learned features, $\sigma$ is the activation function, $\textbf{W}^{(l)}_{\text{G}}$ and $\textbf{W}^{(l)}_{\text{E}}$ are learnable weights for the $l$-th layer, and $\textbf{B}^{(l)}_{\text{G}}$ and $\textbf{B}^{(l)}_{\text{I}}$ are the biases for the $l$-th layer. The features in both modality are initialized as $\textbf{Z}^{(0)}_{\text{I}} = \textbf{X}_{\text{I}}$ and $\textbf{Z}^{(0)}_{\text{G}} = \textbf{X}_{\text{G}}$. The normalized adjacency matrix for $\textbf{A}$ is defined as $\Tilde{\textbf{A}}=\textbf{D}^{-\frac{1}{2}}\textbf{A}\textbf{D}^{\frac{1}{2}}$, where $\textbf{D}$ is the degree matrix of the graph.

The learned features from both GAEs are aggregated using weight at each layer. 
\begin{equation}
    \textbf{Z}^{(l)} = \alpha_l\textbf{Z}^{(l)}_{\text{G}} + (1-\alpha_l)\textbf{Z}^{(l)}_{\text{I}},
\end{equation}
where $\alpha_l$ is a learnable weight for $l$-th layer, and $L$ is total number of GCN layers. The aggregated feature, $Z^{(L)}$ is used as the final learned feature from MPGA and serves as the input for the graph decoder. A graph decoder is used to reconstruct the gene expression data,
\begin{equation}
    \textbf{X}_{\text{Rec}}=\sigma(\Tilde{\textbf{A}}_{\text{G}} \textbf{Z}^{(L)}\textbf{W}_{\text{Rec}}+\textbf{B}_{\text{Rec}}).
\end{equation}
And the reconstruction loss is defined as the follow.
\begin{equation}
    L_{Rec}=\sum_{i=1}^N\|\bm{x}^{G}_i-\bm{x}_i^{\text{Rec}}\|^2,
\end{equation}
 where $\bm{x}_i^{G} \in \mathbb{R}^M$ is the gene expression data for $i$-th spot and $\bm{x}_i^{\text{Rec}} \in \mathbb{R}^M$ is the reconstructed gene expression data for $i$-th spot.

\begin{figure*}[htp]
\begin{center}
\centerline{\includegraphics[width=\textwidth]{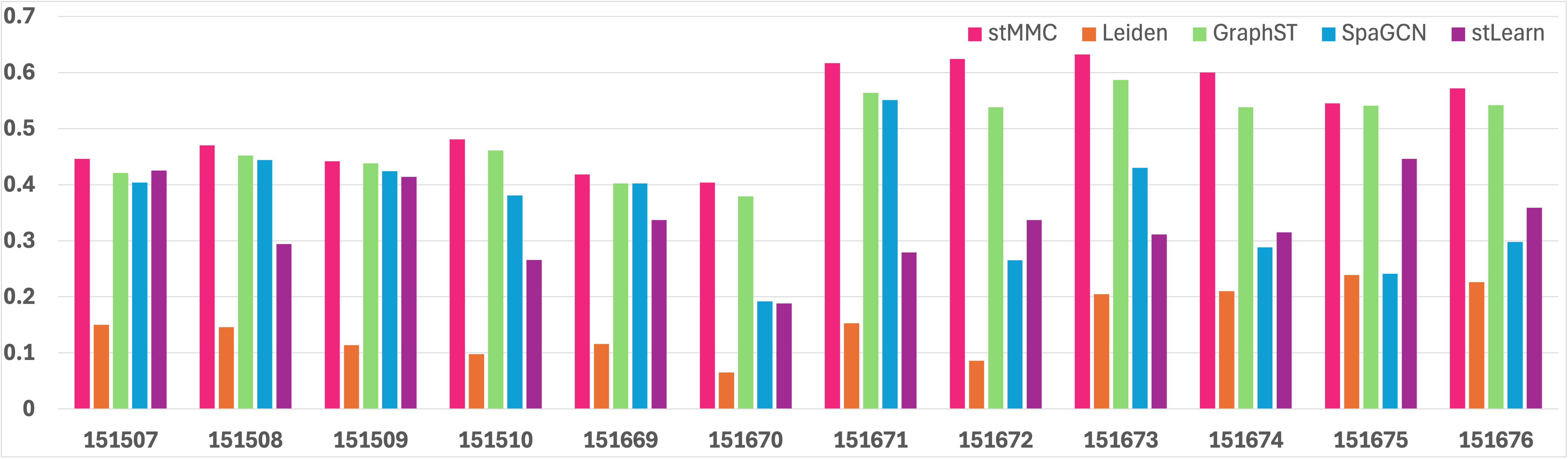}}
\caption{The ARI scores of stMMC and all four baseline models on DLPFC datasets are plotted here, where the $Y$ axis is the ARI score and the $X$ axis is the data slice number.}
\label{fig:DLPFC_ARI}
\end{center}
\vspace{-15pt}
\end{figure*}

\begin{figure*}[htp]
\begin{center}
\centerline{\includegraphics[width=\textwidth]{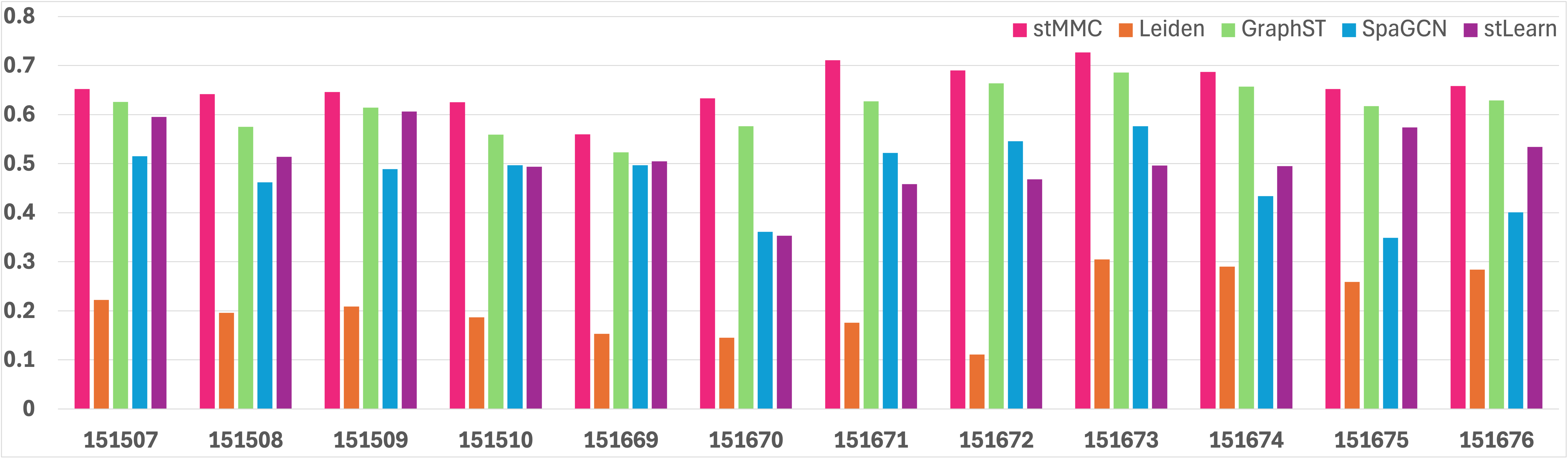}}
\caption{The NMI scores of stMMC and all four baseline models on DLPFC datasets are plotted here, where the $Y$ axis is the NMI score and the $X$ axis is the data slice number.}
\label{fig:DLPFC_NMI}
\end{center}
\vspace{-15pt}
\end{figure*}

\begin{figure*}[h!]
    \centering
    \begin{minipage}[b]{0.3\textwidth}
        \centering
        \includegraphics[width=\textwidth]{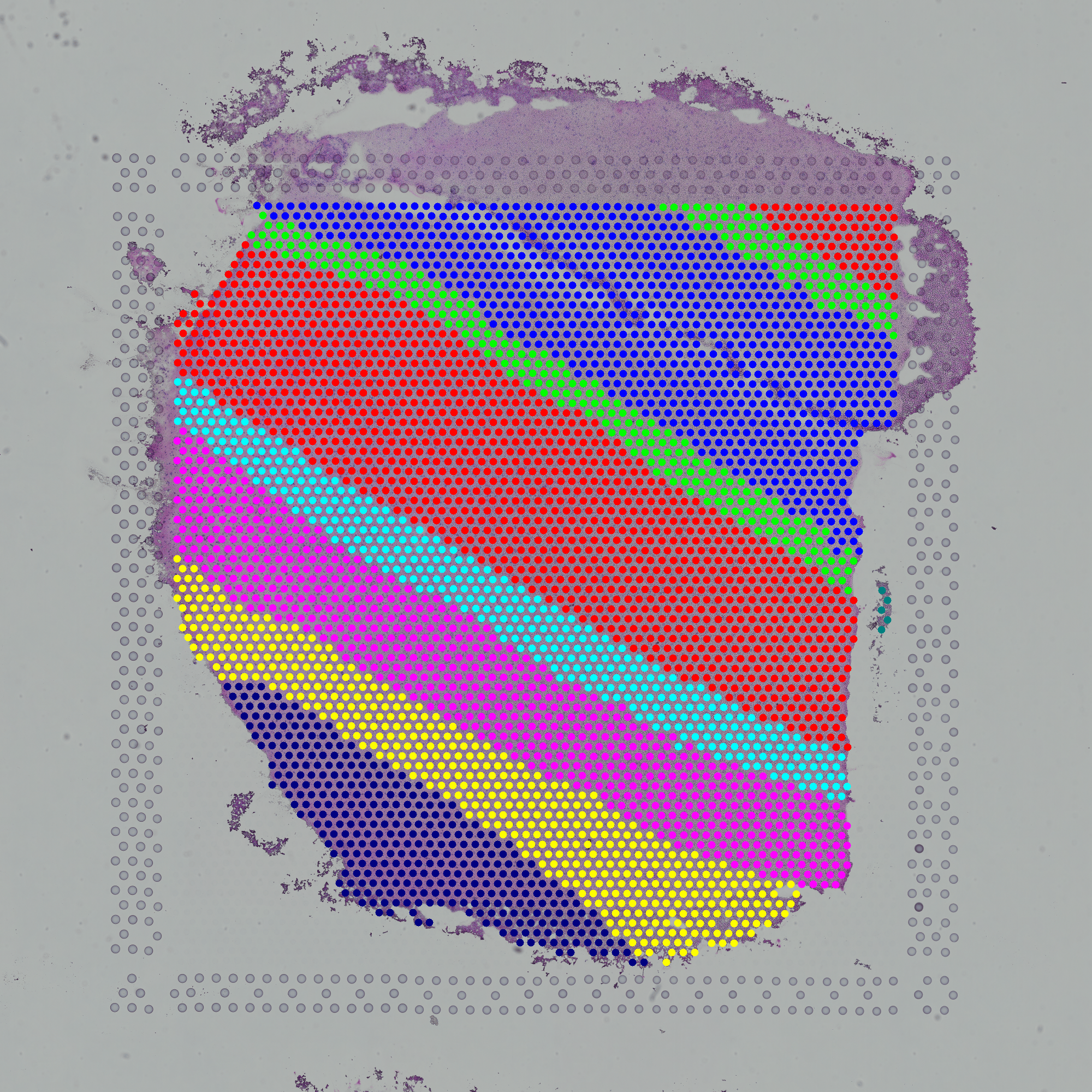}
        \subcaption{Ground Truth}
        \label{fig:dlpfc_gt}
    \end{minipage}
    \hfill
    \begin{minipage}[b]{0.3\textwidth}
        \centering
        \includegraphics[width=\textwidth]{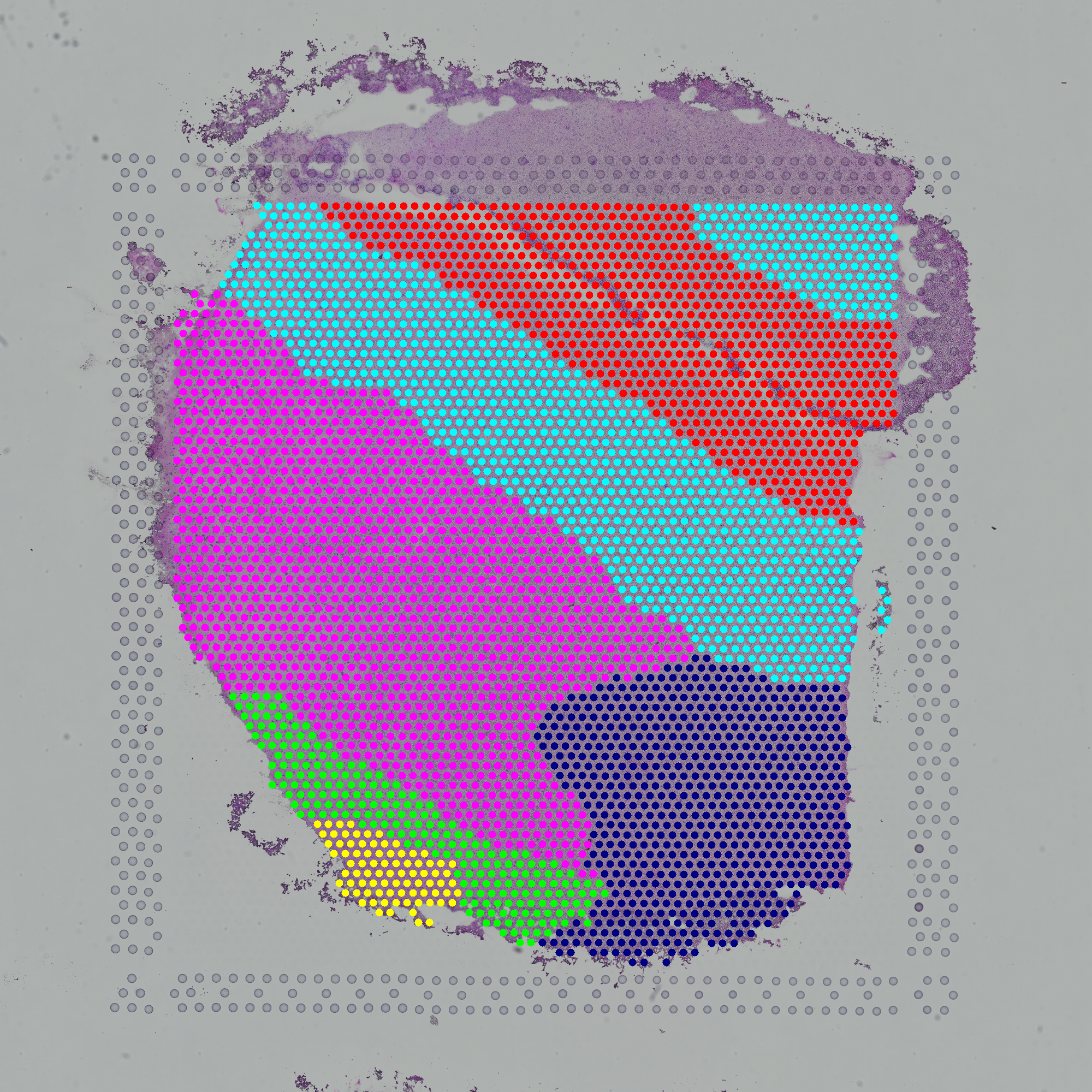}
        \subcaption{stMMC}
        \label{fig:dlpfc_stmmc}
    \end{minipage}
    \hfill
    \begin{minipage}[b]{0.3\textwidth}
        \centering
        \includegraphics[width=\textwidth]{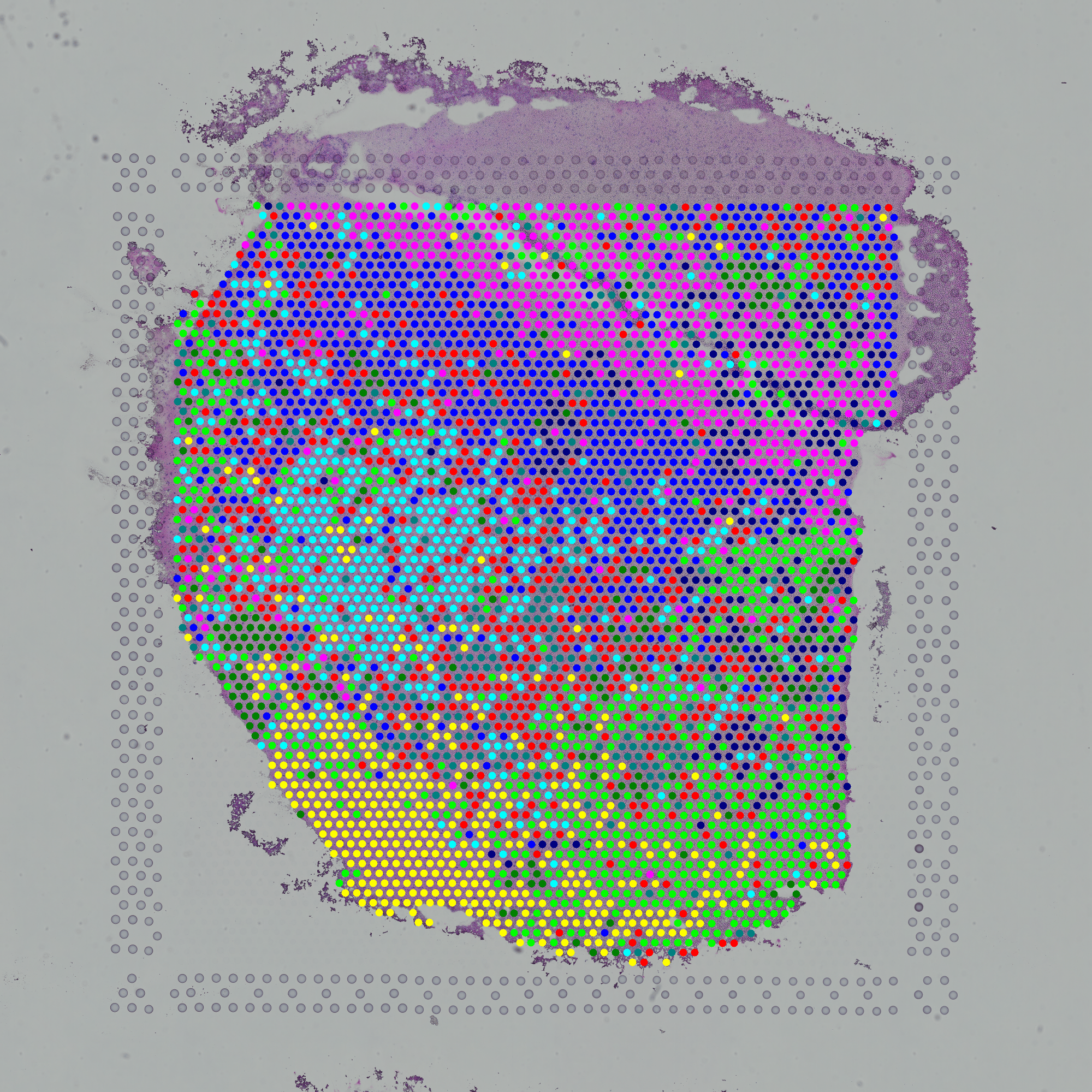}
        \subcaption{Leiden}
        \label{fig:dlpfc_leiden}
    \end{minipage}

    \begin{minipage}[b]{0.3\textwidth}
        \centering
        \includegraphics[width=\textwidth]{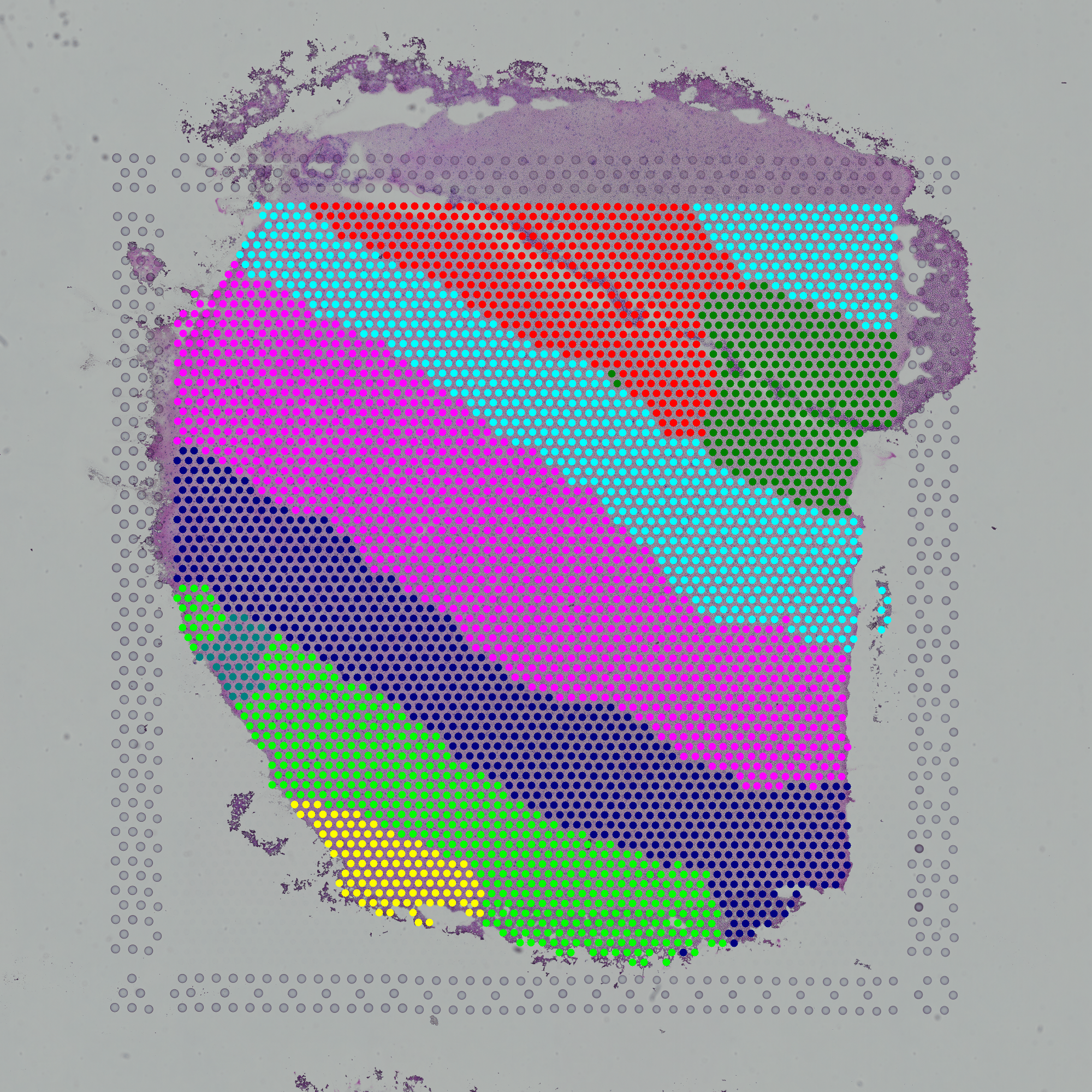}
        \subcaption{GraphST}
        \label{fig:dlpfc_graphst}
    \end{minipage}
    \hfill
    \begin{minipage}[b]{0.3\textwidth}
        \centering
        \includegraphics[width=\textwidth]{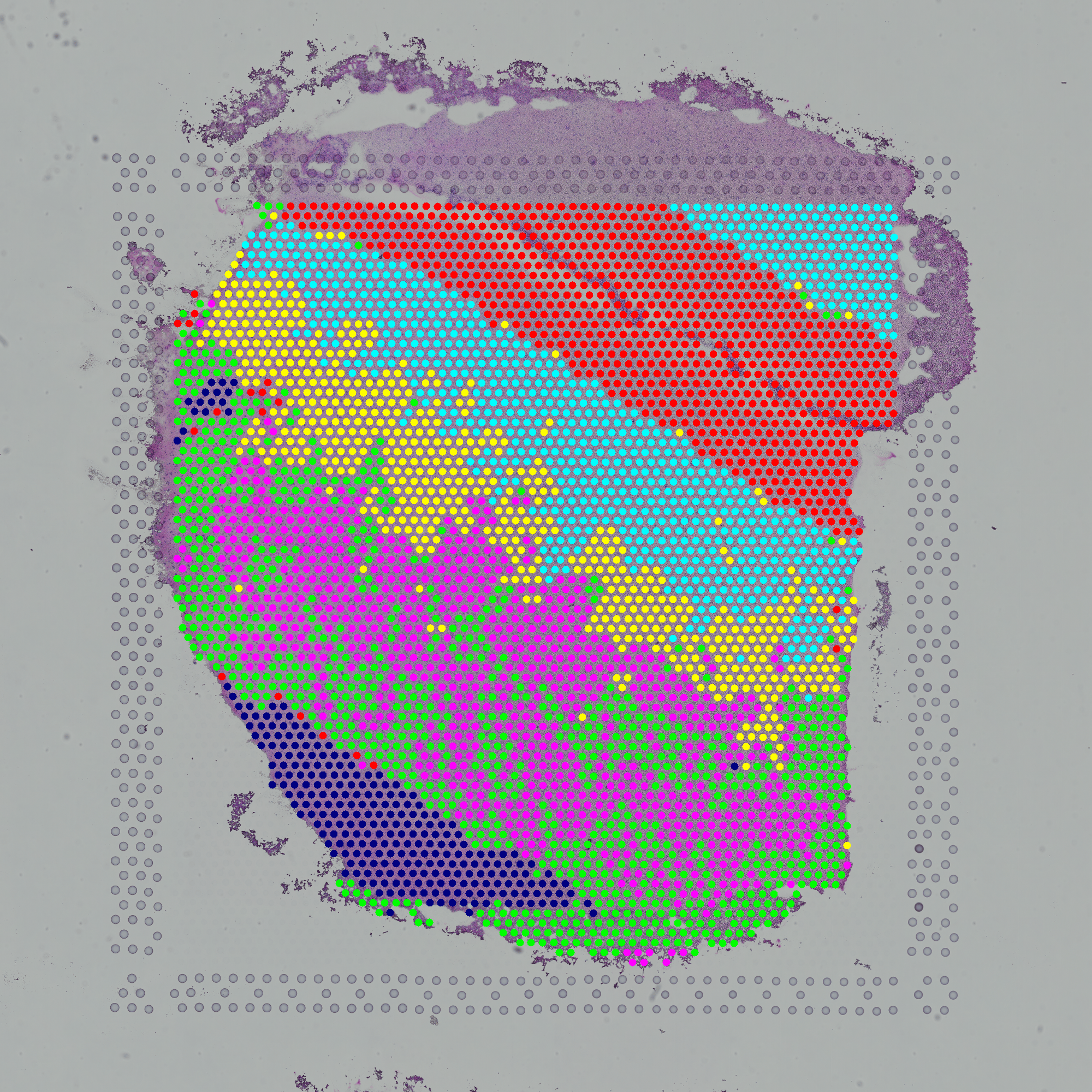}
        \subcaption{SpaGCN}
        \label{fig:dlpfc_spagcn}
    \end{minipage}
    \hfill
    \begin{minipage}[b]{0.3\textwidth}
        \centering
        \includegraphics[width=\textwidth]{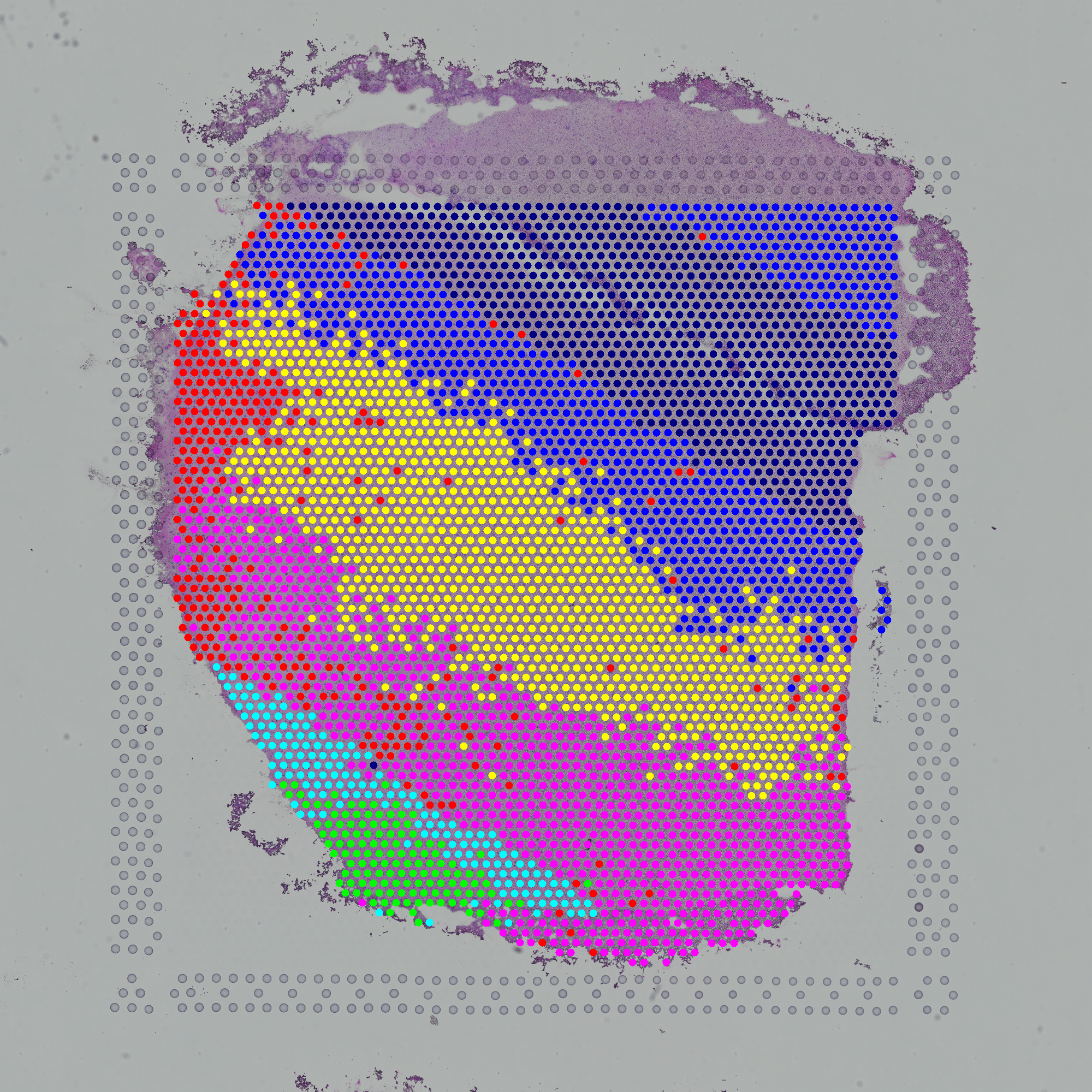}
        \subcaption{stLearn}
        \label{fig:dlpfc_stlearn}
    \end{minipage}
    \caption{All the clustering assignments of the proposed method, stMMC and four baseline models are plotted against the ground truth for DLPFC Slice 151507.}
    \label{fig:dlpfc_visual}
\vspace{-15pt}
\end{figure*}

\subsection{Contrastive Learning Module}
Contrastive learning is an emerging technique used to better extract embedding features in unsupervised learning problems and has shown promising results~\cite{Dong_Pachade_Liang_Sheth_Giancardo_2024,li2023contrastive,li_scgemoc_2023}. The core mechanism of contrastive learning is to assign positive and negative pairs to different feature embeddings, and pull the embeddings with positive pairs close and push the embeddings with negative pairs far away. Inspired by the Deep Graph Infomax approach\cite{velivckovic2018deep}, a corrupted graph is generated for each modality by shuffling nodes while maintaining the same graph topology, denoted as $\mathcal{G}^*_{\text{G}}=(\bm{X}^*_{\text{G}},V^*_{\text{S}},E_{\text{Proximity}})$ and $\mathcal{G}^*_{\text{I}}=(\bm{X}^*_{\text{I}},V^*_{\text{S}},E_{\text{Similarity}})$ for gene expression data and image feature data, respectively. $E_{\text{Proximity}}$ and $E_{\text{Similarity}}$ stay the same during the shuffle. Corrupted graphs are fed into the same GAE within the same modality and corrupted learned features are obtained, denoted as $Z^*_{\text{I}}$ and $Z^*_{\text{G}}$. To obtain the localized community information among spots, a community representation is computed for each spot with the following definition:
\begin{equation}
    \bm{g}^{(m)}_i = \frac{1}{|\text{Neb}(i)|}\sum_{j \in \text{Neb}(i)} \bm{z}_{j,m}, \forall \; m \in [\text{I}, \text{G}]
\end{equation}
where $\bm{z}_{j,m}$ is the learned representation for spot $j$ in $m$-th modality, and $\text{Neb}(i)$ is the set of one-step neighbors of spot $i$. The learned embedding from the original graph, $\bm{z}_{i,m}$ and the community representation from the original graph, $\bm{g}_{i,m}$ are assigned positive pairs, while the learned embedding from the corrupted graph, $z\bm{^}*_{i,m}$ and original community representation, $\bm{g}_i$ are assigned negative pairs. The key idea of the implemented contrastive learning mechanism is that the local community representation of spot $i$, $\bm{g}_{i,m}$ should be close to the original learned embedding of the same spot, $\bm{z}_{i,m}$ in the latent space, but far away from the corrupted learned embedding of the same spot, $\bm{z}^*_{i,m}$. An illustration of this process for a random spot is shown in Figure \ref{fig:contrastive_learning}.

A neural network-based discriminator $\Theta(\cdot)$ is used to distinguish between positive and negative pairs. $\Theta(\bm{z}_{i,m},\bm{g}_{i,m})$ calculates a scalar probability score of the pair $(\bm{z}_{i,m},\bm{g}_{i,m})$ being positive. The contrastive learning loss is defined based on binary cross-entropy loss as follows:

\begin{multline}
    L_{\text{CL}} = -\frac{1}{N} \left( \sum_{m \in [\text{G},\text{I}]} \left( \sum_{i=1}^N \left( \mathbb{E}_{(\textbf{X}_m,\textbf{A}_m)} \left[\log \Theta(\bm{z}_{i,m},\bm{g}_{i,m})\right]\right.\right.\right. \\
    \left.\left.\left.+\mathbb{E}_{(\textbf{X}^*_m,\textbf{A}_m)}\left[\log(1-\Theta(\bm{z}^*_{i,m},\bm{g}_{i,m}))\right] \right) \right) \right).
\end{multline}
As the corrupted graph shares the same topology with the original graph, a symmetric contrastive learning loss is defined for the corrupted graph, $\mathcal{G}^*_m$ to make the overall model more stable.

\begin{multline}
    L_{\text{CL\_C}} = -\frac{1}{N} \left( \sum_{m \in [\text{G},\text{I}]} \right.\\
    \left( \sum_{i=1}^N \left( \mathbb{E}_{(\textbf{X}^*_m,\textbf{A}_m)} \left[\log \Theta(\bm{z}^*_{i,m},\bm{g}^*_{i,m})\right]\right.\right. \\
    \left.\left.+ \mathbb{E}_{(\textbf{X}^*_m,\textbf{A}_m)}\left[\log(1-\Theta(\bm{z}_{i,m},\bm{g}^*_{i,m}))\right] \right) \right),
\end{multline}
where $\bm{g}^*_{i,m}$ is the community representation of spot $i$ on the corrupted graph. The final total loss function is defined as follows:
\begin{equation}
    L_{Total} = \theta_1 L_{\text{Rec}} + \theta_2 (L_{\text{CL}} + L_{\text{CL\_C}}),
\end{equation}
where $\theta_1$ and $\theta_2$ are hyperparameters for the strength of different loss terms. The default values for $\theta_1$ and $\theta_2$ in stMMC are 10 and 1. 

\subsection{Clustering Module}
The reconstructed data is used for spatial clustering through a separate clustering module. The default clustering algorithm used in stMMC is mclust \cite{scrucca2023model}. Other conventional clustering methods can also be applied. We observed that some spots were clustered out of sync with their local spot neighborhood in stMMC's clustering assignments, which led to a deterioration in clustering performance, especially on manually annotated datasets. To address this issue, we introduced an optional smoothing step. After the clustering module produces the initial assignments, each spot is reassigned to the same cluster as the majority cluster of its nearest $b$ neighbors. The optimal value of $b$ is set to 50.

\begin{figure*}[h!]
    \centering
    \begin{minipage}[b]{0.3\textwidth}
        \centering
        \includegraphics[width=\textwidth]{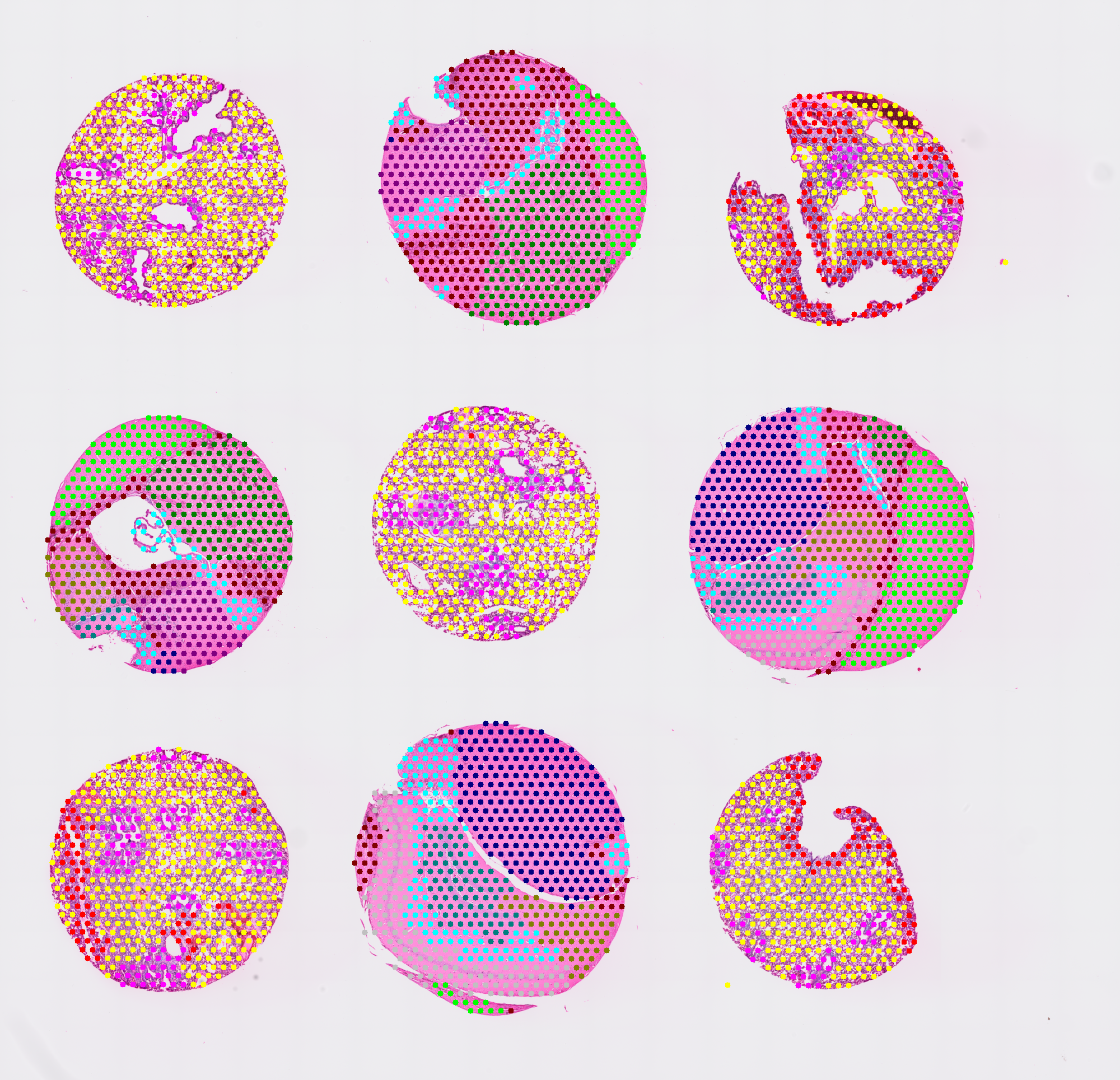}
        \subcaption{Ground Truth}
        \label{fig:mouse_gt}
    \end{minipage}
    \hfill
    \begin{minipage}[b]{0.3\textwidth}
        \centering
        \includegraphics[width=\textwidth]{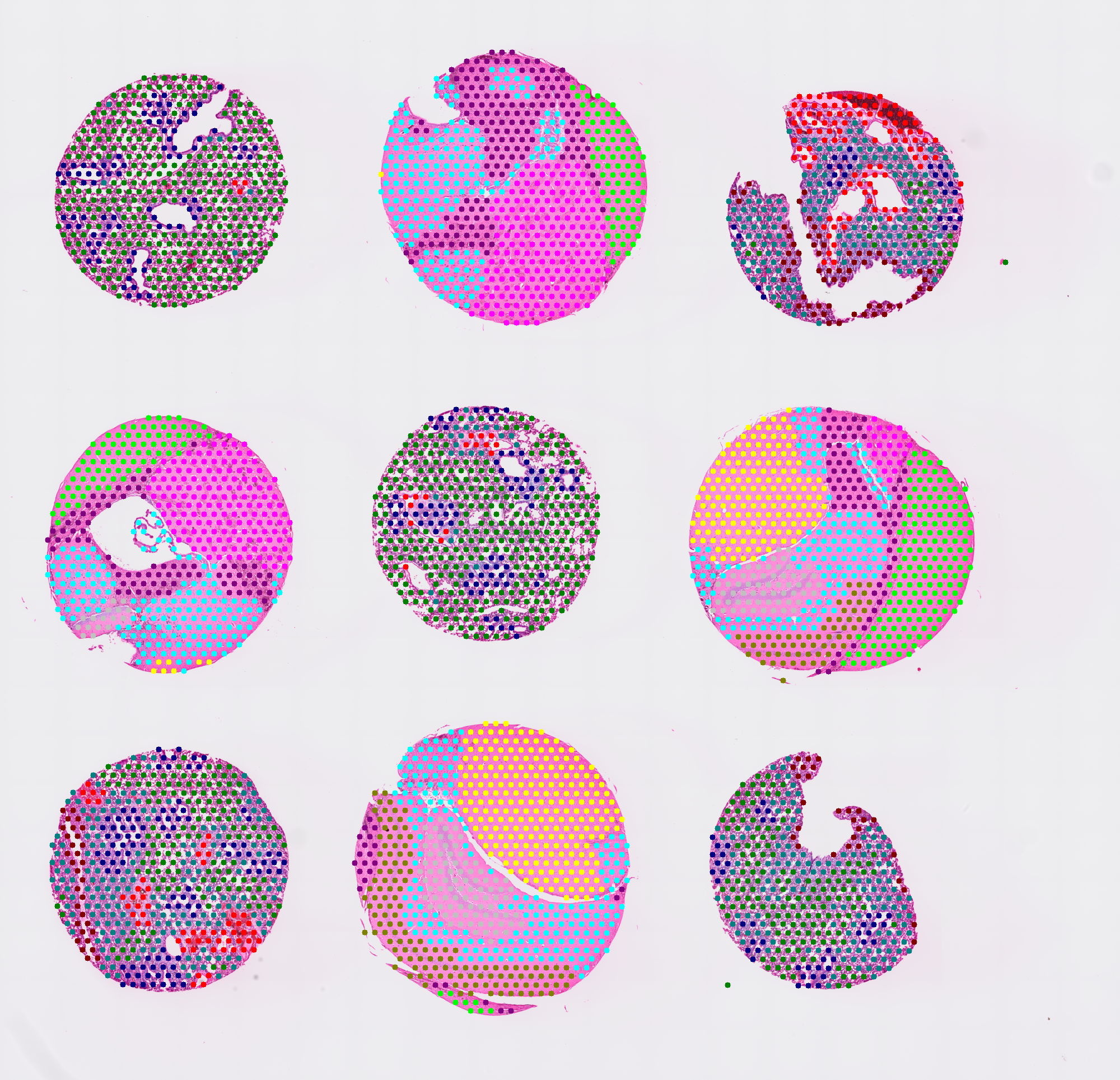}
        \subcaption{stMMC}
        \label{fig:mouse_stmmc}
    \end{minipage}
    \hfill
    \begin{minipage}[b]{0.3\textwidth}
        \centering
        \includegraphics[width=\textwidth]{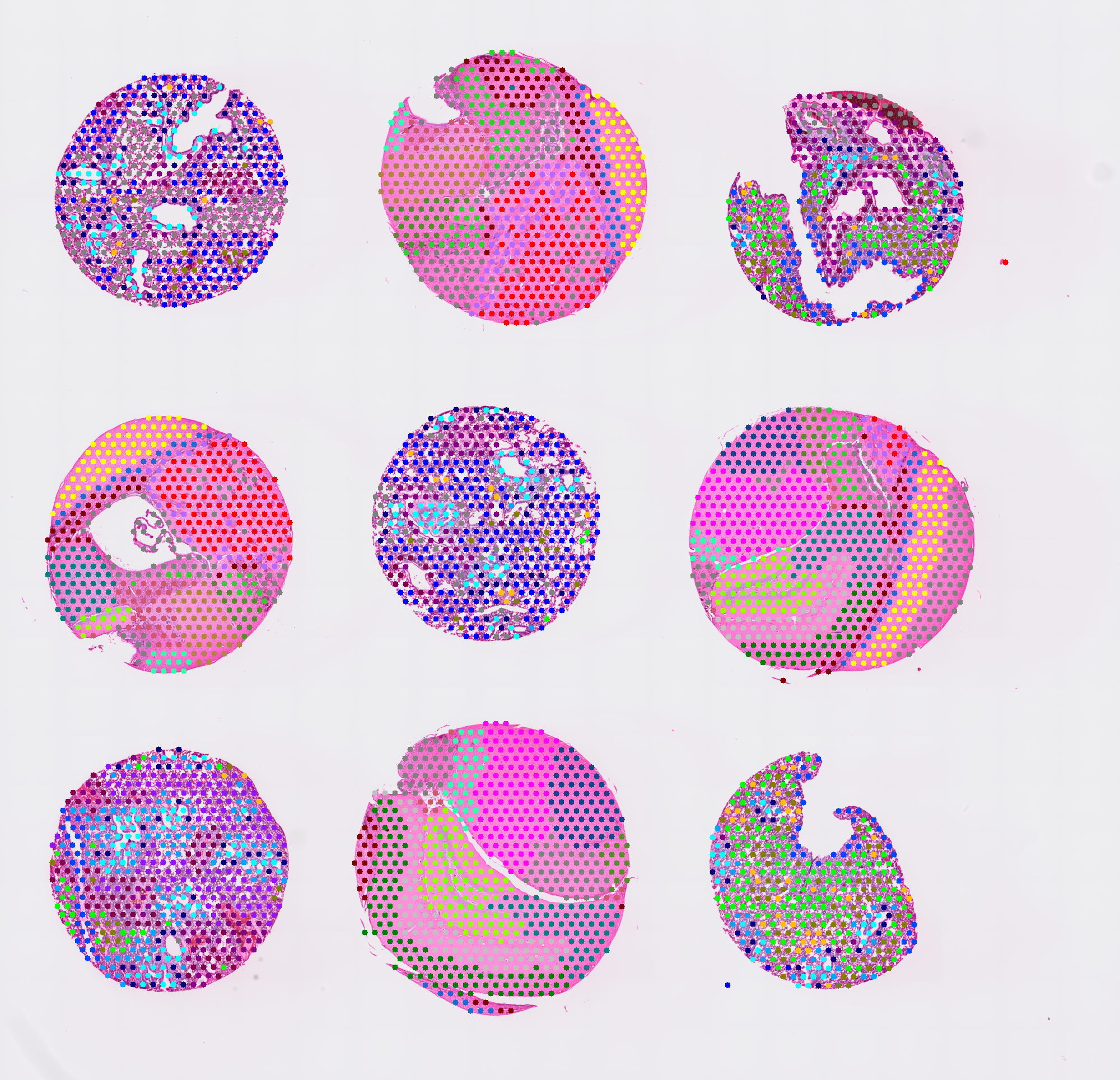}
        \subcaption{Leiden}
        \label{fig:mouse_leiden}
    \end{minipage}

    \begin{minipage}[b]{0.3\textwidth}
        \centering
        \includegraphics[width=\textwidth]{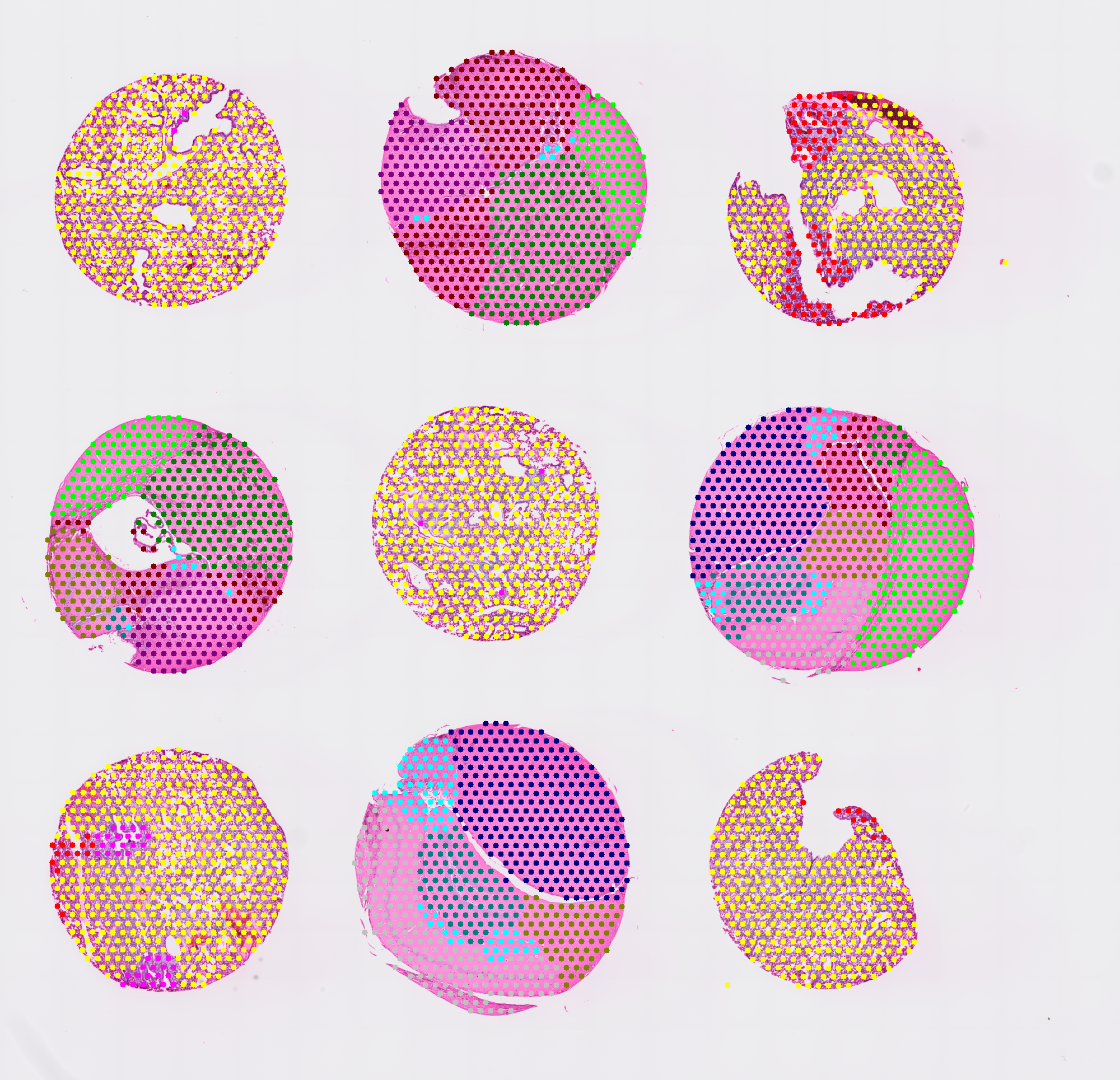}
        \subcaption{GraphST}
        \label{fig:mouse_graphst}
    \end{minipage}
    \hfill
    \begin{minipage}[b]{0.3\textwidth}
        \centering
        \includegraphics[width=\textwidth]{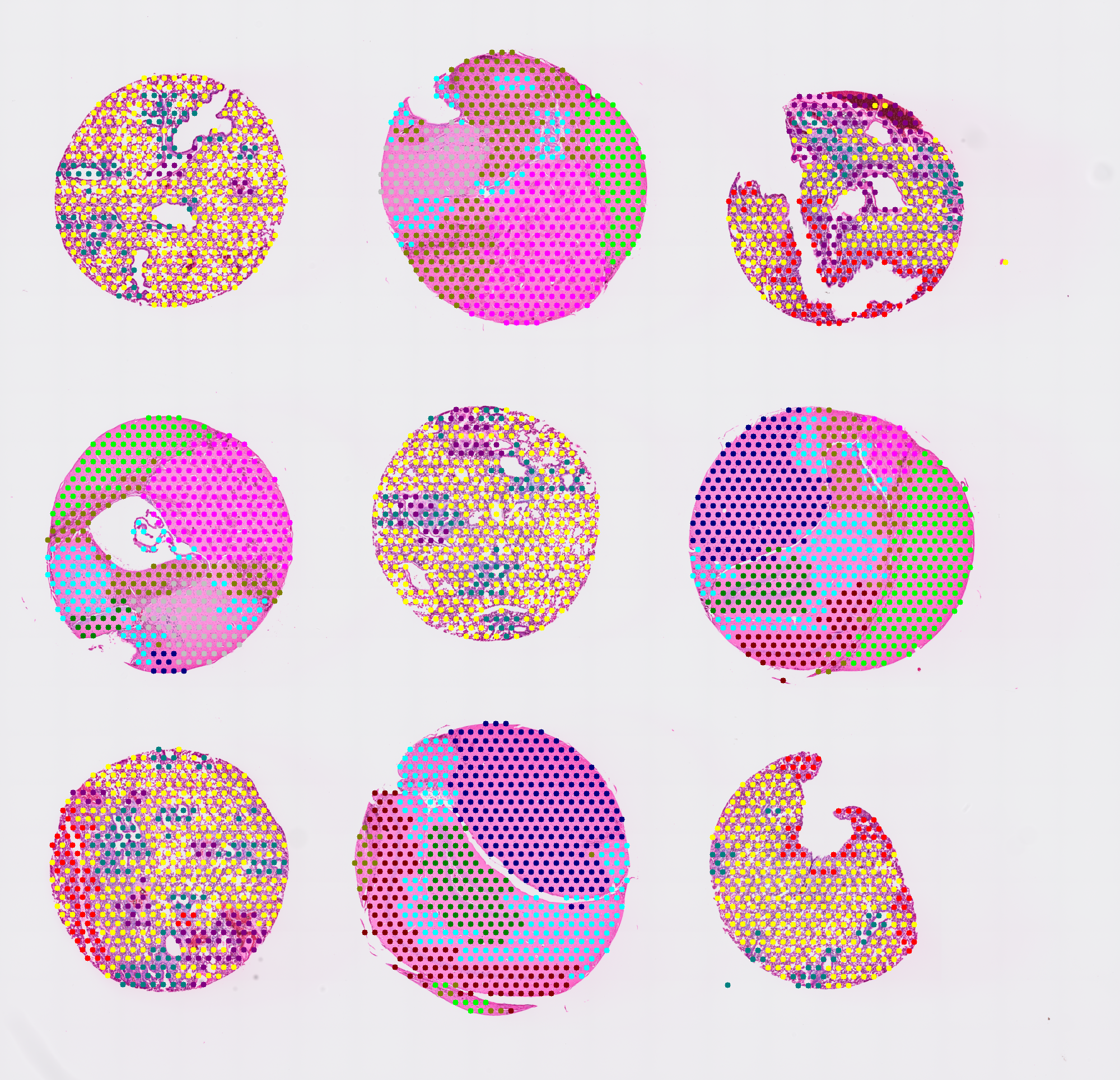}
        \subcaption{SpaGCN}
        \label{fig:mouse_spagcn}
    \end{minipage}
    \hfill
    \begin{minipage}[b]{0.3\textwidth}
        \centering
        \includegraphics[width=\textwidth]{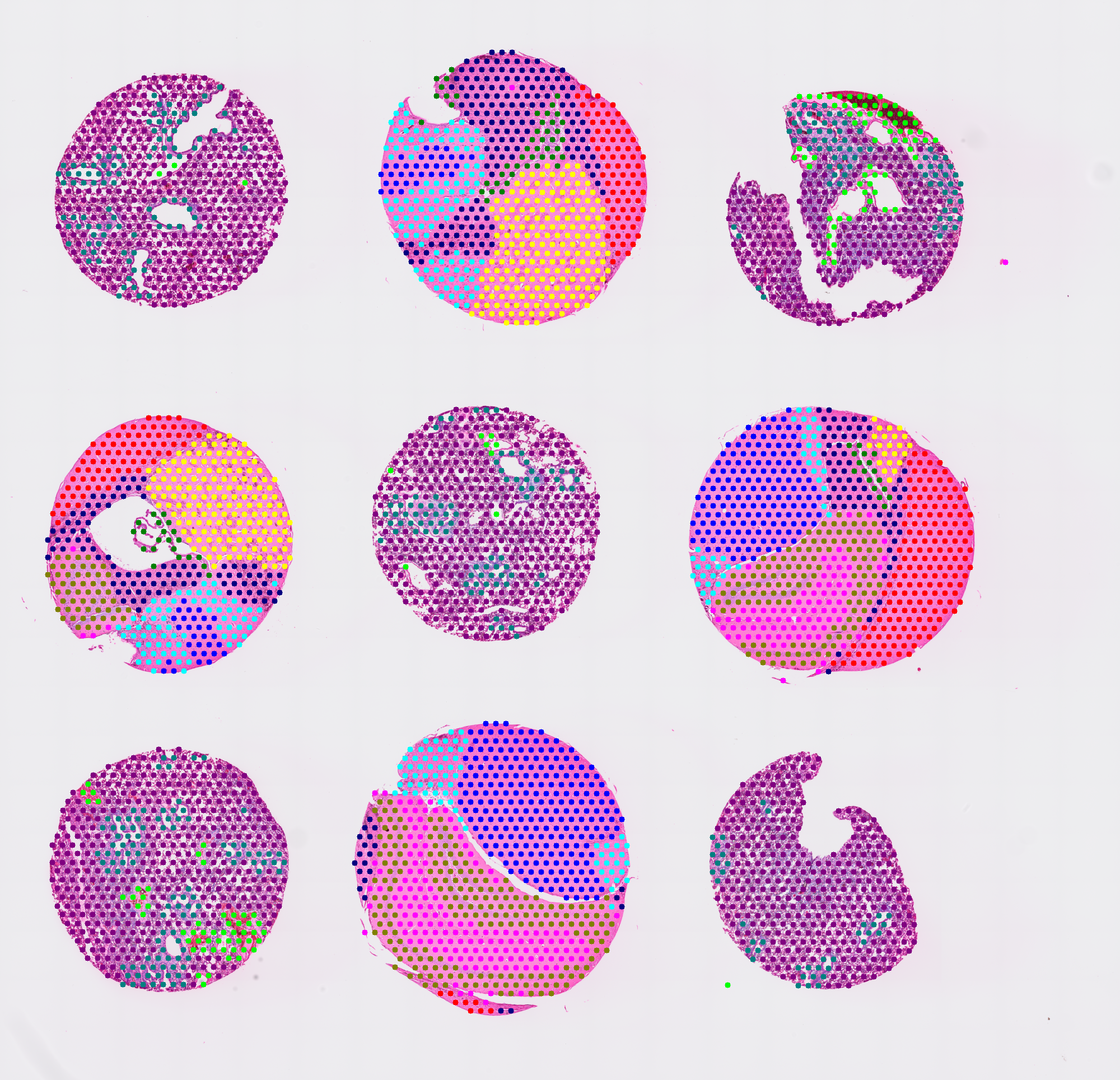}
        \subcaption{stLearn}
        \label{fig:mouse_stlearn}
    \end{minipage}
    \caption{All the clustering assignments of the proposed method, stMMC and four baseline models are plotted against the ground truth for the Mouse 3x3 1mm dataset.}
    \label{fig:mouse_visual}
\vspace{-15pt}
\end{figure*}

\section{Results}
\subsection{Experiment Settings}
We tested stMMC against four state-of-the-art baseline models that cover all three of the major categories, Leiden, GraphST, stLearn, and SpaGCN. 
\begin{itemize}
    \item \textbf{Leiden}: a widely used conventional clustering method for graph structured data~\cite{traag2019louvain}.
    \item \textbf{GraphST}: a GNN-based method that uses contrastive learning to combine location information with gene expression data~\cite{long_spatially_2023}.
    \item \textbf{SpaGCN}: a GCN-based method that combines location information, gene expression data, and a three-dimensional RGB score from the histology image for clustering~\cite{hu_spagcn_2021}.
    \item \textbf{stLearn}: a hybrid method that combines conventional methods and deep learning network extracted morphological image features to obtain spatial morphological gene expression-normalized data for clustering~\cite{pham_robust_2023}.
\end{itemize}

We conducted the experiments on two datasets containing 13 slice samples in total:  i) DLPFC (LIBD human dorsolateral pre-frontal cortex)~\cite{maynard2021transcriptome}, and ii) mouse\_3x3\_1mm from 10x Genomics~\cite{10xgenomics_mouse_tissue_microarray}. The DLPFC dataset is a widely used benchmark dataset for spatial transcriptomics and is also the only one of the two datasets annotated by human experts. The dataset, mouse\_3x3\_1mm is annotated by a graph-based clustering model in 10x Genomics Space Ranger v2.0.1. However, the histology image in mouse\_3x3\_1mm has a much higher resolution than that in DLPFC. While both datasets are collected using 10x genomics Visium platform with the diameter of the spot at about 55 $\mu m$, the image patches of a spot from DLPFC are about 96 pixels in width and the image patches from mouse\_3x3\_1mm are about 219 pixels in width. The number of clusters ranges from 6 to 9 on DLPFC slices and is 12 on the mouse\_3x3\_1mm dataset. stMMC is implemented in Python and is available at \url{https://github.com/NabaviLab/stMMC}. Adjusted random index (ARI) and normalized mutual information (NMI) are the metrics used in this experiment. All experiments are conducted on a server with an 18-core CPU and 2 Nvidia RTX A5000 GPUs.

\subsection{Results on Human Annotated Data}
The performance of stMMC and all four baseline models in terms of ARI and NMI is shown in Figure \ref{fig:DLPFC_ARI} and Figure \ref{fig:DLPFC_NMI}, respectively. It can be observed that stMMC clearly outperforms almost all four baseline models in both metrics. GraphST shows consistent performance as the second-best model across almost all slices, which is in line with the previous benchmark paper~\cite{yuan_benchmarking_2024}. stLearn demonstrates inconsistent performance, achieving second or a very close third-best performance in about half of the sample slices, but also performing close to the worst model in some slices. The performance of SpaGCN is also inconsistent across slices but mostly worse than stLearn. Leiden performs the worst among all models as expected as it is the only model that is not designed specifically for ST data.

To better understand the clustering performance beyond the two metrics, the actual clustering assignments of stMMC and baseline models for slice 151507 are mapped back onto the histology image shown in Figure \ref{fig:dlpfc_visual}. Compared to all baseline models, stMMC is the best at capturing major segmentation between clusters and its separation of the cluster is closer to the ground truth. GraphST captures some major segmentation but also splits larger clusters where no actual segmentation exists. Both SpaGCN and stLearn correctly capture the segmentation of major clusters in the top-right corner, but both suffer from out-of-sync small clusters.

\begin{table}[ht]
\centering
\caption{ARI and NMI Scores of stMMC and Baseline Models on mouse\_3x3\_1mm}
\resizebox{0.5\columnwidth}{!}{
\begin{tabular}{l|ll}
\hline
Model   & ARI & NMI \\
\hline
stMMC   & \textbf{0.527} & \textbf{0.761} \\
Leiden  &  0.398 & 0.687 \\
GraphST & 0.459 & 0.680 \\
SpaGCN  & 0.431 & 0.636 \\
stLearn & 0.511 & 0.760 \\
\hline
\end{tabular}
}
\label{tab:mouse}
\end{table}

\subsection{Results on Model Annotated Data}
For the mouse\_3x3\_1mm dataset, the performance of stMMC and all four baseline models in terms of ARI and NMI is shown in Table \ref{tab:mouse}. The proposed method, stMMC clearly outperforms all four baseline models. We also think the higher resolution histology image in mouse\_3x3\_1mm helps stMMC achieve a better overall performance in terms of ARI and NMI. The clustering assignments of stMMC and four baseline models are mapped back onto the original histology image as shown in Figure \ref{fig:mouse_visual}. The segmentation by stMMC and stLearn is very close to the ground truth, while other baseline models split some of the major clusters. As stLearn is the only other model that utilizes the histology image features, this result shows the benefit of incorporating histology images in spatial clustering.

\begin{table}[ht]
\centering
\caption{ARI and NMI Scores of stMMC's Variations on DLPFC Slice 151673}
\resizebox{\columnwidth}{!}{
\begin{tabular}{lll|ll}
\hline
\multicolumn{3}{c|}{Components}                                                                      & \multirow{2}{*}{ARI}   & \multirow{2}{*}{NMI}   \\ \cline{1-3}
Smoothing & Contrastive & Image &       &       \\
\hline
\cmark & \cmark & \cmark & 0.632 & 0.727 \\
\xmark & \cmark & \cmark & 0.582 & 0.679 \\
\cmark & \xmark & \cmark & 0.586 & 0.668 \\
\cmark & \cmark & \xmark & 0.603 & 0.683 \\
\xmark & \xmark & \cmark & 0.556 & 0.617 \\
\cmark & \xmark & \xmark & 0.577 & 0.635 \\
\xmark & \cmark & \xmark & 0.571 & 0.631 \\
\xmark & \xmark & \xmark & 0.523 & 0.594 \\
\hline
\end{tabular}
}
\label{tab:abaltion}
\end{table}

\subsection{Ablation Study}
To determine the contributions of each major component in stMMC, an ablation study on the contrastive learning mechanism, image feature modality, and the smoothing step is conducted on DLPFC dataset slice 151673 as shown in Table \ref{tab:abaltion}.
For variations that omit one component, stMMC without contrastive learning mechanism and stMMC without smoothing step causes the worst performance deterioration, with over 7\% drop in ARI and about 8\% drop in NMI, and stMMC without image feature modality causes a slightly smaller drop in performance, 4.6\% drop in ARI and 6.1\% drop in NMI. For variations that omit two components, stMMC without both the contrastive learning mechanism and smoothing step sees a worse performance than the other two. And stMMC without all three components is the worst performing one as expected. We can conclude that the contrastive learning mechanism and smoothing steps are the two most impactful components in stMMC and all the major components contribute as expected.

\section{Conclusions}
In this study, we proposed stMMC, a novel deep-learning spatial clustering method for spatial transcriptomics, which leverages a contrastive learning mechanism for better feature extraction. To the best of our knowledge, stMMC is the first method that integrates histology image features as an additional modality for spatial clustering. stMMC effectively integrates gene expression data and histology image features by using a multi-modal parallel graph autoencoder, which is proven to improve the clustering performance over state-of-the-art baseline models in experiments.

We conducted experiments with stMMC and four baseline models on two public datasets consisting of 13 sample slices in total, DLPFC and mouse\_3x3\_1mm. The experimental results demonstrate that stMMC consistently outperforms state-of-the-art baseline methods in terms of ARI and NMI scores. Visualization of the clustering assignments shows that stMMC is better at capturing important cluster separations. Furthermore, the ablation study validates the contributions of the three major components, the contrastive learning mechanism and the incorporation of the histology image features. These results indicate that stMMC is not only spatial clustering in spatial transcriptomics but also sets a foundation for future works that could further explore clustering in spatial transcriptomics as multi-modal data.

\section*{Acknowledgment}
This work is supported by the National Science Foundation (NSF) under grant No. 1942303, PI: Nabavi and grant No. 2348278, PI: Nabavi.

\bibliographystyle{IEEEtran}
\bibliography{reference}

 \end{document}